\documentclass{article}

\usepackage[preprint]{neurips_2026}

\usepackage[utf8]{inputenc} 
\usepackage[T1]{fontenc}    
\usepackage{hyperref}       
\usepackage{url}            
\usepackage{booktabs}       
\usepackage{amsfonts}       
\usepackage{nicefrac}       
\usepackage{microtype}      
\usepackage{xcolor}         
\usepackage{booktabs}
\usepackage{pifont}
\newcommand{\cmark}{\ding{51}}
\newcommand{\xmark}{\ding{55}}
\usepackage{graphicx}   
\usepackage{array}
\usepackage{makecell}   
\usepackage{natbib}
\usepackage{amsmath}
\usepackage{amssymb}
\usepackage{booktabs}
\usepackage{graphicx}  
\usepackage{makecell}   
\usepackage{caption}
\usepackage{subcaption}
\usepackage{sidecap}

\usepackage[most]{tcolorbox}
\usepackage{listings}

\usepackage{xcolor}

\definecolor{promptgreenbg}{RGB}{245,250,246}
\definecolor{promptgreenframe}{RGB}{46,125,50}
\definecolor{promptgreentitle}{RGB}{232,245,233}
\definecolor{promptgreenrule}{RGB}{27,94,32}

\newtcolorbox{promptbox}[1][]{
  enhanced,
  breakable,
  colback=promptgreenbg,
  colframe=promptgreenframe,
  colbacktitle=promptgreentitle,
  coltitle=black,
  boxrule=0.9pt,
  arc=2pt,
  left=8pt,
  right=8pt,
  top=8pt,
  bottom=8pt,
  fonttitle=\bfseries,
  title={Pairwise LLM-Judge Prompt},
  borderline west={1.5pt}{0pt}{promptgreenrule},
  #1
}

\newtcolorbox{extractionbox}[1][]{
  enhanced,
  breakable,
  colback=blue!3!white,
  colframe=blue!45!black,
  colbacktitle=blue!8!white,
  coltitle=black,
  boxrule=0.8pt,
  arc=2pt,
  left=6pt,
  right=6pt,
  top=6pt,
  bottom=6pt,
  fonttitle=\bfseries,
  title={Example of benchmark extraction from a source paper},
  #1
}
\newcommand{\rcf}{\textsc{rcf}}
\newcommand{\hpa}{\textsc{hpa}}
\newcommand{\msi}{\textsc{msi}}
\newcommand{\sns}{\textsc{sns}}
\newcommand{\ip}{\textsc{ip}}
\newcommand{\pdq}{\textsc{pdq}}
\newcommand{\cbs}{\textsc{cbs}}
\newcommand{\bench}{\textsc{Matter to Mechanism}}

\title{Matter to Mechanism: A Benchmark for AI Co-Scientists in Materials and Battery Research}

\author{%
Shashwat Sourav$^{1,2,4,5*}$ \And
Tanjin He$^{3,4*}$ \And
Maria K. Y. Chan$^{3,4\dagger}$ \And
Anubhav Jain$^{4\dagger}$ \And
Tirthankar Ghosal$^{2,\dagger}$ \\[0.5em]
$^{1}$Washington University in St. Louis \\
$^{2}$Oak Ridge National Laboratory \\
$^{3}$Argonne National Laboratory \\
$^{4}$Lawrence Berkeley National Laboratory \\
$^{5}$UniverseTBD \\[0.5em]
\texttt{\{shashwat.sourav\}@wustl.edu} \\
\texttt{\{tanjin.he,mchan\}@anl.gov \quad ajain@lbl.gov} \\
\texttt{ghosalt@ornl.gov} \\[0.3em]
\small $^{*}$Equal contribution.
}

\begin{document}

\maketitle

\begin{abstract}

AI co-scientists are increasingly used for scientific discovery, but current evaluations still do not test them on a key task: moving from a concrete scientific or technological problem to a plausible, mechanism-grounded solution hypothesis. This gap is especially important in materials science and, in particular, battery research, where a useful proposal must identify the relevant failure mode, propose a credible intervention, and explain why that intervention should improve the target property. We introduce Matter to Mechanism, a benchmark for evaluating AI co-scientists on problem-to-hypothesis reasoning in materials science, with a focus on battery materials research. The benchmark contains 2,645 instances derived from scientific publications. Each instance includes a structured problem statement, a candidate solution hypothesis, an explicit reasoning trace, and domain-grounded annotations such as material system, component, failure mode, intervention, mechanism, target property, and claimed outcome. We also introduce a metric suite that measures reasoning fidelity, problem alignment, mechanistic specificity, novelty, plausibility, and problem decomposition quality, and combine them into a composite score. Using this framework, we evaluate several AI co-scientist systems and show that Matter to Mechanism reveals interpretable system differences that are only partially recovered by standard text-similarity metrics. We further show through adversarial stress tests that the aggregate score is more stable than individual metric dimensions under superficial gaming attacks.
\end{abstract}

\section{Introduction}

Large language models are now being used in more and more parts of scientific research. They are being asked to read papers, summarize prior work, retrieve useful evidence, write code, analyze data, and, in some cases, even propose new ideas. This has created excitement around the idea of AI co-scientists that can assist human researchers during discovery. At the same time, evaluation is still lagging behind the ambition of these systems. In many cases, we still do not have good ways to test whether a model can perform the scientific tasks of interest, rather than simply produce fluent scientific text or score well on a narrow benchmark \citep{2024arXiv240701725P,2025arXiv250321248L,2025arXiv250300096M,2024arXiv241005080C}.

This gap is observed in materials science. Materials science research is not just a text generation problem. A useful scientific proposal has to begin from an observation, e.g. a problem with the performance of a functional material, identify what is failing, suggest an appropriate intervention, and hypothesize why that intervention should improve the target property. In battery research, for example, a strong hypothesis should do more than mention a material, additive, or interface. It should connect the proposed change to a specific mechanism such as dendrite suppression, interfacial stabilization, ionic transport improvement, structural robustness, or degradation mitigation. A proposal can sound scientific and still be vague, misaligned with the actual problem, or physically implausible. That is exactly why this setting needs a benchmark that goes beyond surface similarity or general human preference judgments  \citep{2021arXiv210707002D,2021arXiv211115366R,2025arXiv250510852L}.

Recent work has started to push benchmark design in the right direction. DiscoveryBench studies data-driven scientific discovery and evaluates whether models can derive hypotheses and workflows from datasets and discovery goals \citep{2024arXiv240701725P}. ResearchBench decomposes scientific discovery into inspiration retrieval, hypothesis composition, and hypothesis ranking, and shows that these sub-tasks can be benchmarked separately \citep{2025arXiv250321248L}. BixBench evaluates LLM-based agents on realistic computational biology analysis scenarios, while ScienceAgentBench focuses on data-driven scientific workflows and code-based agent performance \citep{2025arXiv250300096M,2024arXiv241005080C}. In materials and chemistry, MATDESIGN studies hypothesis generation for material discovery and design, MatTools benchmarks LLMs on materials-science tool use, MaCBench evaluates multimodal chemistry and materials reasoning, and ChemBench evaluates chemical knowledge and reasoning at scale \citep{2025arXiv250113299K,2025arXiv250510852L,2024arXiv241116955A,2024arXiv240401475M}. These are important steps, but they still leave open a specific problem that matters in practice (Table~\ref{tab:benchmark_comparison}): evaluating whether a system can start from a concrete materials-science problem, identify the relevant failure mode, propose an intervention at the right level, and connect that intervention to a scientifically plausible mechanism. This problem is narrower than general scientific-discovery benchmarking, but also closer to the kind of mechanistically grounded reasoning that matters in real materials and battery research.

In this work, we introduce \bench{}, a benchmark for evaluating AI co-scientists on problem-solution reasoning in materials science, in particular battery materials research. Each instance is built from a scientific paper and contains a problem statement, a candidate hypothesis, and an explicit reasoning trace. In addition, each instance includes structured scientific fields such as the material or battery system, battery component, failure mode, intervention, mechanism, target property, claimed outcome, evidence strength, and novelty axis. This structure is important because it turns the benchmark into more than a collection of prompts and answers. It gives us a way to ask whether the model identified the right problem, proposed the right kind of intervention, and linked that intervention to a mechanism that makes scientific sense.

We also introduce a metric suite designed for this setting. Instead of depending only on lexical overlap or broad quality judgments \citep{2021arXiv210707002D,2021arXiv211115366R}, we evaluate systems along six dimensions: reasoning chain fidelity, hypothesis-problem alignment, mechanistic specificity, scientific novelty, intervention plausibility, and problem decomposition quality. These dimensions are meant to reflect how a domain researcher would distinguish a genuinely useful hypothesis from one that is linguistically polished but scientifically superficial. The resulting benchmark therefore plays two roles. First, it provides a structured dataset of materials-science problem-solution instances. Second, it provides an evaluation framework for understanding what current AI co-scientists can and cannot do in scientific discovery.

\begin{table*}[t]
\centering
\scriptsize
\setlength{\tabcolsep}{3.2pt}
\renewcommand{\arraystretch}{1.1}
\resizebox{\textwidth}{!}{%
\begin{tabular}{lcccccc}
\toprule
\textbf{Benchmark} &
\makecell{\textbf{Problem}\\\textbf{$\rightarrow$ Hypothesis}} &
\makecell{\textbf{Materials /}\\\textbf{Battery}} &
\makecell{\textbf{Structured}\\\textbf{Problem Decomp.}} &
\makecell{\textbf{Reasoning}\\\textbf{Trace}} &
\makecell{\textbf{Domain}\\\textbf{Metrics}} &
\makecell{\textbf{LLM /}\\\textbf{Agent Eval.}} \\
\midrule
DiscoveryBench \citep{2024arXiv240701725P} & \textcolor{green}{\cmark} & \textcolor{red}{\xmark} & \textcolor{red}{\xmark} & \textcolor{red}{\xmark} & \textcolor{red}{\xmark} & \textcolor{green}{\cmark} \\
ResearchBench \citep{2025arXiv250321248L} & \textcolor{green}{\cmark} & \textcolor{red}{\xmark} & \textcolor{green}{\cmark} & \textcolor{red}{\xmark} & \textcolor{red}{\xmark} & \textcolor{green}{\cmark} \\
BixBench \citep{2025arXiv250300096M} & \textcolor{red}{\xmark} & \textcolor{red}{\xmark} & \textcolor{red}{\xmark} & \textcolor{red}{\xmark} & \textcolor{red}{\xmark} & \textcolor{green}{\cmark} \\
MATDESIGN \citep{2025arXiv250113299K} & \textcolor{green}{\cmark} & \textcolor{green}{\cmark} & \textcolor{red}{\xmark} & \textcolor{red}{\xmark} & \textcolor{green}{\cmark} & \textcolor{green}{\cmark} \\
MatTools \citep{2025arXiv250510852L} & \textcolor{red}{\xmark} & \textcolor{green}{\cmark} & \textcolor{red}{\xmark} & \textcolor{red}{\xmark} & \textcolor{red}{\xmark} & \textcolor{green}{\cmark} \\
MaCBench \citep{2024arXiv241116955A} & \textcolor{red}{\xmark} & \textcolor{green}{\cmark} & \textcolor{red}{\xmark} & \textcolor{red}{\xmark} & \textcolor{red}{\xmark} & \textcolor{green}{\cmark} \\
ChemBench \citep{2024arXiv240401475M} & \textcolor{red}{\xmark} & \textcolor{red}{\xmark} &
\textcolor{red}{\xmark} & \textcolor{red}{\xmark} & \textcolor{green}{\cmark} & \textcolor{green}{\cmark} \\
\textbf{Matter to Mechanism} & \textcolor{green}{\cmark} & \textcolor{green}{\cmark} & \textcolor{green}{\cmark} & \textcolor{green}{\cmark} & \textcolor{green}{\cmark} & \textcolor{green}{\cmark} \\
\bottomrule
\end{tabular}%
}
\caption{
Comparison of Matter to Mechanism with representative benchmarks for scientific discovery, materials science, and LLM/agent evaluation. Existing benchmarks evaluate important aspects of discovery, including data-driven hypothesis generation, inspiration-based decomposition, multimodal materials reasoning, and materials tool use. Our benchmark differs in jointly focusing on problem--solution reasoning in materials science, structured scientific decomposition, explicit reasoning traces, and domain-grounded evaluation metrics.
}
\label{tab:benchmark_comparison}
\end{table*}

Our goal is not only to compare different co-scientists, but also to highlight their shortcomings. A model may generate a hypothesis that is relevant but does not actually address the stated failure mode. It may propose something novel but without a clear mechanism. It may reason for many steps without decomposing the scientific problem well enough to support a useful intervention. By making these distinctions visible, Matter to Mechanism shifts evaluation away from ``does the output sound scientific?'' and toward ``does the system perform the scientific task well?'' Overall, our contribution is a benchmark and evaluation framework for a part of scientific discovery that is important, practical, and still poorly measured. We hope this benchmark helps move the field toward more realistic evaluation of AI systems for materials research, especially in high-value domains such as batteries where the importance of mechanistic grounding and physical plausibility stands out.

\begin{figure}[t]
    \centering
    \includegraphics[width=\linewidth]{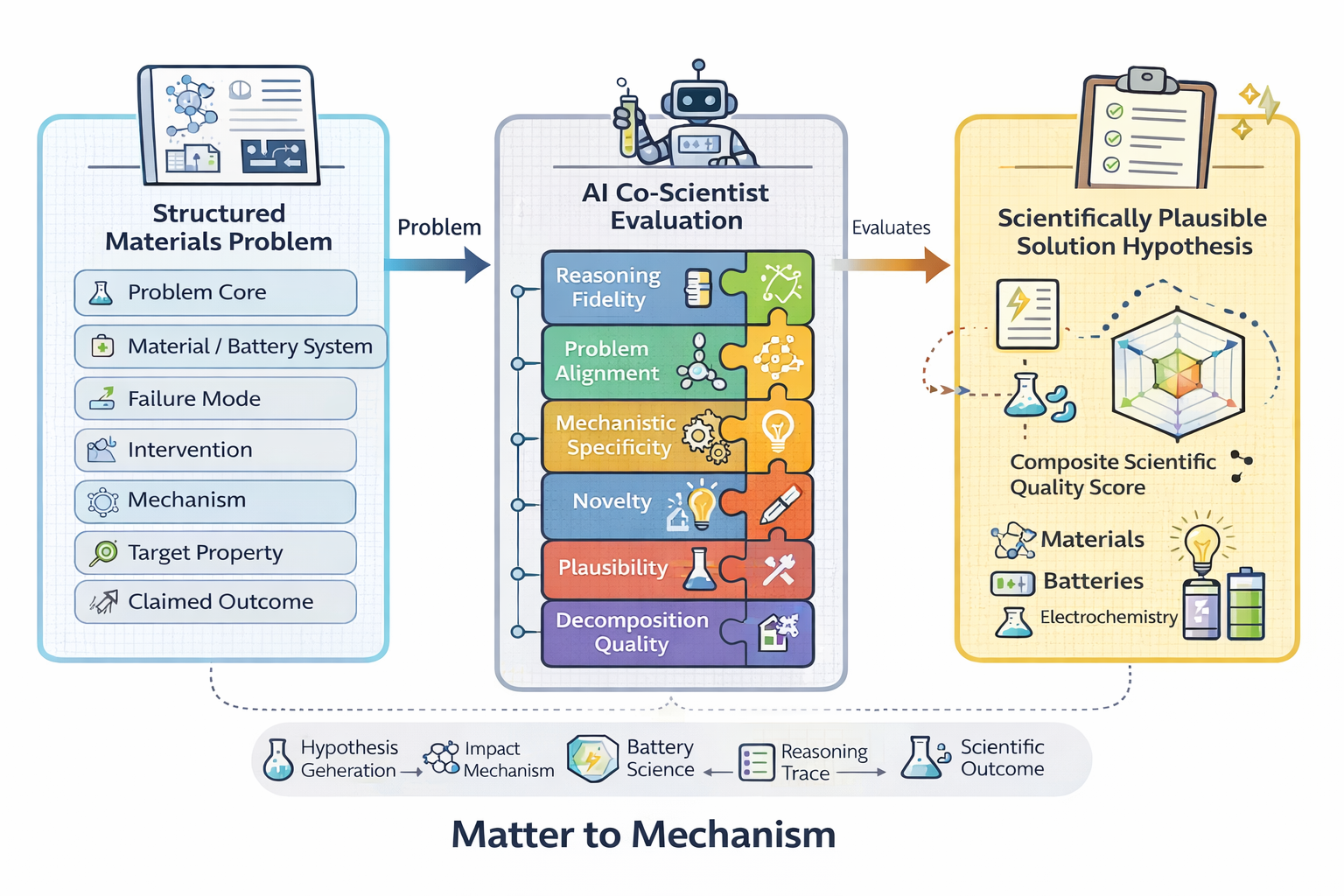}
    \caption{
    Overview of the Matter to Mechanism workflow. Given a structured materials-science problem, AI co-scientist systems generate candidate solution hypotheses and associated reasoning traces. These outputs are evaluated along six dimensions: reasoning fidelity, problem alignment, mechanistic specificity, novelty, plausibility, and decomposition quality, which together yield a composite scientific quality score. The framework is designed to assess whether models can move from problem formulation to mechanistically grounded and scientifically plausible hypotheses.
    }
    \label{fig:1}
\end{figure}

\section{Dataset Description}
\label{sec:dataset}

\bench{} focuses on problem-solution reasoning, starting from a materials problem, identifying what is failing, and proposing a solution that is relevant, mechanistically specific, and physically plausible. Unlike benchmarks that evaluate broad discovery workflows or general hypothesis ranking, our dataset represents each example as a structured mapping from a problem-side description to a solution-side description.

\paragraph{Task formulation.}

We represent each example as a structured mapping from a problem description to a solution description. The problem description includes the problem statement, the material or battery system, the affected component, and the failure mode. The solution description includes the hypothesis, the intervention, the mechanism, the target property, the claimed outcome, and the reasoning trace. Under this formulation, the benchmark asks whether a system can move from what is failing to what should be changed and why.

\paragraph{Dataset construction.}

We built the benchmark from a corpus of 90{,}000 open-access papers. We filter this corpus for papers relevant to materials science and battery research and retain 2 {,}645 papers. Each paper is converted into one structured problem-solution instance linking a technical problem to a candidate solution hypothesis and an associated reasoning trace. The extraction pipeline identifies problem-side fields, such as the problem statement, material or battery system, and failure mode, as well as solution-side fields, such as mechanism, target property, claimed outcome, and evidence strength. The goal of this processing step is to preserve the scientific logic of the source paper while converting it into a format suitable for evaluation. Two domain experts reviewed the schema, and field definitions.

\begin{tcolorbox}[colback=blue!5!white, colframe=blue!75!black, title=\textbf{Example of benchmark extraction from a source paper}]
\small
\textbf{Source signal.} A representative paper studies a disordered-rocksalt cathode for lithium-ion batteries and identifies restricted Li$^+$ diffusion, low electrical conductivity, and irreversible oxygen loss as bottlenecks for capacity retention.

\medskip
\textbf{Extracted problem fields.}
Battery system: lithium-ion battery.
Component: disordered-rocksalt cathode.
Failure mode: restricted Li$^+$ diffusion, low conductivity, and oxygen loss.
Problem core: poor capacity retention caused by transport and redox-instability bottlenecks.

\medskip
\textbf{Extracted solution fields.}
Intervention: fluorination of the oxygen sublattice, particle-size reduction by ball milling, and carbon coating.
Mechanism: fluorination improves Li$^+$ percolation and suppresses irreversible oxygen loss; ball milling shortens diffusion paths; carbon coating improves electronic transport.
Target property: Li$^+$ diffusion kinetics, charge-transfer resistance, and oxygen-redox reversibility.
Claimed outcome: improved discharge capacity, rate capability, and cycling stability.

\medskip
\textbf{Reasoning trace.} The reasoning trace links the identified bottleneck to the proposed intervention and then to a mechanism: transport limitations motivate fluorination and particle-size reduction, electronic limitations motivate carbon coating, and the combined intervention is expected to improve both kinetics and structural reversibility.

\end{tcolorbox}

\textbf{Reasoning trace.} The reasoning trace links the identified bottleneck to the proposed intervention and then to a mechanism: transport limitations motivate fluorination and particle-size reduction, electronic limitations motivate carbon coating, and the combined intervention is expected to improve both kinetics and structural reversibility.

\paragraph{Multiple plausible hypotheses.}
A given materials problem can often admit more than one scientifically plausible solution, and different reasoning traces may support different interventions. For this reason, Matter to Mechanism does not treat the literature-derived hypothesis as the only valid answer. Instead, the literature instance serves as an anchor for benchmark construction, while our main evaluation remains reference-free with respect to hypothesis matching. Generated outputs are scored primarily against the input problem fields and their own internal scientific structure, rather than by exact overlap with a single reference hypothesis. This design makes the benchmark more suitable for open-ended scientific reasoning, where multiple solutions may exist.

\paragraph{Benchmark overview.}

We present Matter to Mechanism (Figure~\ref{fig:1}), a benchmark for evaluating AI co-scientists on problem-solution reasoning in materials science, with a focus on battery discovery. Each example is derived from a scientific paper and includes a problem statement, a candidate hypothesis, and a reasoning trace. Each example also includes structured scientific fields such as the material system, component, failure mode, intervention, mechanism, target property, claimed outcome, evidence strength, and novelty axis. This structure makes the benchmark useful for evaluation because it lets us measure not only whether an output is on topic but also whether it addresses the right scientific bottleneck in the right way. 

\paragraph{Data source and composition.}

The benchmark contains 2{,}645 problem--solution instances collected from materials-science papers, with strong coverage of battery-related research. All core structured fields are present in the released benchmark, and each example contains a reasoning trace. The benchmark covers a wide range of battery systems, components, and problem types, including transport limitations, degradation, safety, interfacial instability, diagnostics, characterization, and thermal-management challenges. This breadth is important because it reduces the risk that evaluation collapses into narrow paraphrase matching within one chemistry or one failure mode.

\paragraph{Evaluation.}

This benchmark is intended to support the evaluation of AI co-scientists under a shared and realistic task formulation. Recent works have emphasized that evaluation itself should be studied carefully, with attention to benchmark design, scope, and interpretation \citep{2024arXiv240713168T,2023arXiv231008677Z,2024arXiv241024100L}. Prior work has shown that benchmark results can be fragile and can overstate what is actually being measured \citep{2021arXiv210707002D,2021arXiv211115366R,2012arXiv1206.4656W}. Hence, our goal is to provide a benchmark that measures a specific, high-value task in scientific reasoning and makes system failures more interpretable.

\paragraph{Metric-aware design.}

The benchmark is paired with a domain-based evaluation framework. We score outputs along six dimensions: reasoning-chain fidelity, hypothesis-problem alignment, mechanistic specificity, scientific novelty, intervention plausibility, and problem decomposition quality (details are given in Appendix~\ref{app:metrics}). We treat these metrics as proxies for different aspects of hypothesis quality that domain researchers care about, such as whether the output addresses the stated failure mode, proposes an intervention at the right level, connects that intervention to a specific mechanism, remains scientifically feasible, and decomposes the problem enough to support solution building. This design is intended to separate outputs that only sound scientific from outputs that address the right problem with the right mechanism. We also report a composite score, but we treat the individual dimensions as important in because they highlight different failure modes, and we use our validation experiments (Section \ref{sec:experiments}) to test whether the aggregate captures signal beyond conventional metrics and remains more stable under adversarial manipulation.

\paragraph{Contribution.}

Matter to Mechanism contributes both a dataset and an evaluation framework. It provides structured problem-solution instances grounded in published materials-science papers. It also provides a way to study whether current co-scientists can move from a concrete scientific problem to a plausible and mechanistically grounded solution. We view this as a targeted benchmark for one important aspect of scientific discovery, rather than as a claim to measure discovery in full.

\section{Experiments}
\label{sec:experiments}

We use \bench{} to study how well current AI systems can map a structured scientific problem to a plausible solution hypothesis. Our experiments are guided by three questions. First, can current systems approach the quality of literature-derived hypotheses under a domain-grounded evaluation? Second, which aspects of scientific reasoning remain difficult? Third, do our validation experiments show that the benchmark captures distinctions that are not fully explained by generic text-similarity metrics?

\paragraph{Models.}

We evaluate ChemDFM-8B as a domain-specialized baseline \citep{2025CRPS....602523Z}; \textsc{Gemini-Weak} and \textsc{Gemini-Direct} as direct-generation baselines with weaker and stronger prompting; \textsc{Gemini-Retrieval} as a retrieval-augmented variant; \textsc{Open Co-Scientist} as an open LangGraph implementation of a Google co-scientist-style workflow inspired by \citep{2025arXiv250218864G}; and \textsc{AI-Researcher} \citep{2025arXiv250518705T} as a general scientific-agent baseline. All systems receive the same problem-side input and are asked to generate a hypothesis together with structured scientific outputs. They are not given the literature hypothesis.

\paragraph{Evaluation Metrics.}

\bench{} evaluates generated hypotheses along six reference-free dimensions, each scaled to $[0,1]$. The goal is not to measure text similarity to a single reference answer, but to measure whether the generated hypothesis addresses the scientific problem in a structured and credible way. This design choice follows the broader view in recent evaluation work that benchmarks should make their target construct explicit and should avoid overstating what a single metric can mean  \citep{2021arXiv210707002D,2021arXiv211115366R,2024arXiv241024100L}. The six dimensions capture different parts of the task. Reasoning Chain Fidelity (\rcf{}) measures whether the reasoning trace is coherent and non-redundant. Hypothesis-Problem Alignment (\hpa{}) measures whether the hypothesis actually addresses the stated problem. Mechanistic Specificity Index (\msi{}) measures whether the output contains concrete mechanistic content rather than generic scientific language. Scientific Novelty Score (\sns{}) measures how distinct the proposal is relative to the benchmark corpus. Intervention Plausibility (\ip{}) measures whether the intervention is physically and materially plausible. Problem Decomposition Quality (\pdq{}) measures whether the original problem has been framed sharply enough to support solution building. We also report a weighted aggregate, \cbs{}, but we do not treat it as a complete substitute for the six individual dimensions. 

\paragraph{Evaluation setup.}

Our main evaluation is reference-free with respect to hypothesis matching: the six core metrics score the generated output against the input problem fields and its own internal structure, rather than against a gold reference answer (Table~\ref{tab:metric_summary}). This choice is motivated by the fact that scientific problems can admit multiple plausible hypotheses, and by prior concerns that benchmark conclusions can become too dependent on narrow matching criteria \citep{2021arXiv210707002D,2021arXiv211115366R}. We report all six metric dimensions together with the composite \cbs{} score.

\paragraph{Validation experiments.}
We also include validation experiments beyond the main leaderboard. We compare \cbs{} against generic reference-based metrics such as BLEU \citep{2018arXiv180408771P,2019arXiv190604903T}, ROUGE-L, BERTScore \citep{2019arXiv190409675Z}, and embedding cosine similarity. We also perform pairwise LLM-judge validation \citep{2024arXiv241115594G} using Gemini-family judges with swap-order evaluation, and we run a metric-gaming stress test using adversarial outputs such as problem mirroring, jargon stuffing, and verbose fake reasoning. These validation experiments help clarify what the benchmark captures, what it does not capture, and how robust it is to simple forms of manipulation.

\paragraph{Weighting scheme}

The \cbs{} weights (Table~\ref{tab:metric_summary}) are intended to balance multiple aspects of scientific usefulness rather than rewarding only fluent text. We assign the largest weights to \rcf{} and \hpa{} because an output that does not address the stated problem, or whose reasoning is internally incoherent, is unlikely to be useful even if it appears scientific. We also weight \msi{} and \ip{} heavily because materials hypotheses must be grounded in a clear mechanism and remain physically plausible. \sns{} and \pdq{} are weighted somewhat lower because novelty and decomposition quality are most meaningful once the output is already relevant, mechanistically specific, and plausible. We therefore treat \cbs{} as a structured summary of the six dimensions rather than a complete measure of scientific quality, and we rely on the validation experiments (Section~\ref{sec:generic_validation}-\ref{sec:gaming_validation}) to test whether this aggregate behaves sensibly in practice. An example of how the weighting scheme works is shown in Appendix \ref{app:worked_example}. 


\begin{table}[t]
\centering
\caption{
Summary of the six core \bench{} metrics. Details of metric computation are provided in Appendix~\ref{app:metrics}.
}
\label{tab:metric_summary}
\small
\resizebox{\columnwidth}{!}{%
\begin{tabular}{llll}
\toprule
\textbf{Metric} & \textbf{Main purpose} & \textbf{Main inputs} & \textbf{Weight} \\
\midrule
\rcf{} & reasoning coherence & reasoning, hypothesis & 0.20 \\
\hpa{} & problem alignment & hypothesis, problem fields & 0.20 \\
\msi{} & mechanistic detail & hypothesis, mechanism, reasoning & 0.18 \\
\sns{} & corpus-level novelty & hypothesis, intervention, system & 0.15 \\
\ip{}  & plausibility & intervention, mechanism, outcome & 0.15 \\
\pdq{} & problem decomposition & problem-side fields & 0.12 \\
\midrule
\cbs{} & weighted aggregate & total & 1.00 \\
\bottomrule
\end{tabular}%
}
\end{table}

\section{Results}
\label{sec:results}

We evaluate six systems on our benchmark. The literature-derived \textsc{Reference} baseline, \textsc{ChemDFM-8B}, \textsc{Gemini-Direct}, \textsc{Gemini-Weak}, \textsc{AI-Researcher}, and \textsc{Open Co-Scientist}. Our goal is not only to compare systems by a single aggregate score, but also to understand how they differ across reasoning quality, problem alignment, mechanistic depth, novelty, plausibility, and problem decomposition. Table~\ref{tab:main_results} reports the main system-level results. The literature-derived reference set remains the strongest overall system, with a \cbs{} of 0.4646. Among generated systems, \textsc{ChemDFM-8B} performs best overall with a \cbs{} of 0.4266, followed by \textsc{Gemini-Direct} (0.3878), \textsc{AI-Researcher} (0.3860), \textsc{Gemini-Weak} (0.3790), and \textsc{Open Co-Scientist} (0.3790). The gap between the reference and the strongest generated system is therefore non-trivial, which is a useful property for the benchmark: it suggests that the benchmark is not saturated and still distinguishes literature-quality hypotheses from current co-scientist outputs. At the same time, the metric breakdown shows that the systems do not fail in the same way. \textsc{Gemini-Direct} and \textsc{AI-Researcher} obtain the highest \rcf{} scores among the generated systems, indicating strong reasoning-chain structure and fluent multi-step outputs. However, both remain very weak on \msi{}, showing that high-quality reasoning traces do not automatically translate into mechanistically specific hypotheses.

\begin{table*}[t]
\centering
\small
\setlength{\tabcolsep}{4pt}
\renewcommand{\arraystretch}{1.1}
\caption{
Main leaderboard on \bench{}. Scores are means over all instances; 95\% bootstrap confidence intervals (2{,}000 resamples) are shown in brackets. The reference corpus remains strongest overall. Among generated systems, \textsc{ChemDFM-8B} achieves the best composite score. A key pattern is that high \rcf{} does not imply high \msi{}: several systems produce coherent reasoning traces while remaining weak on mechanistic specificity.
}
\label{tab:main_results}
\resizebox{\textwidth}{!}{%
\begin{tabular}{lcccccccc}
\toprule
\textbf{System} & \textbf{N} &
\textbf{CBS} & \textbf{RCF} & \textbf{HPA} &
\textbf{MSI} & \textbf{SNS} & \textbf{IP} & \textbf{PDQ} \\
\midrule
\textsc{Reference} & 2645 &
$0.467_{\,[0.465,\,0.469]}$ &
$0.763_{\,[0.758,\,0.768]}$ &
$0.110_{\,[0.106,\,0.114]}$ &
$0.207_{\,[0.201,\,0.212]}$ &
$0.715_{\,[0.711,\,0.719]}$ &
$0.536_{\,[0.530,\,0.541]}$ &
$0.562_{\,[0.556,\,0.567]}$ \\
\midrule
\textsc{ChemDFM-8B} & 2645 &
$0.427_{\,[0.424,\,0.429]}$ &
$0.554_{\,[0.547,\,0.559]}$ &
$0.257_{\,[0.252,\,0.262]}$ &
$0.171_{\,[0.163,\,0.178]}$ &
$0.724_{\,[0.718,\,0.729]}$ &
$0.385_{\,[0.382,\,0.389]}$ &
$0.562_{\,[0.556,\,0.567]}$ \\
\textsc{Gemini-Direct} & 2645 &
$0.388_{\,[0.387,\,0.389]}$ &
$0.746_{\,[0.744,\,0.747]}$ &
$0.189_{\,[0.188,\,0.191]}$ &
$0.038_{\,[0.037,\,0.040]}$ &
$0.724_{\,[0.721,\,0.728]}$ &
$0.504_{\,[0.502,\,0.506]}$ &
$0.501_{\,[0.498,\,0.504]}$ \\
\textsc{Gemini-Retrieval} & 2645 &
$0.388_{\,[0.387,\,0.389]}$ &
$0.745_{\,[0.744,\,0.746]}$ &
$0.188_{\,[0.187,\,0.190]}$ &
$0.040_{\,[0.038,\,0.041]}$ &
$0.724_{\,[0.721,\,0.728]}$ &
$0.504_{\,[0.502,\,0.506]}$ &
$0.501_{\,[0.498,\,0.504]}$ \\
\textsc{AI-Researcher} & 2645 &
$0.386_{\,[0.384,\,0.388]}$ &
$0.740_{\,[0.737,\,0.743]}$ &
$0.188_{\,[0.185,\,0.192]}$ &
$0.039_{\,[0.035,\,0.043]}$ &
$0.692_{\,[0.688,\,0.696]}$ &
$0.504_{\,[0.500,\,0.507]}$ &
$0.498_{\,[0.491,\,0.504]}$ \\
\textsc{Open Co-Scientist} & 2645 &
$0.379_{\,[0.377,\,0.381]}$ &
$0.739_{\,[0.737,\,0.741]}$ &
$0.158_{\,[0.155,\,0.160]}$ &
$0.042_{\,[0.039,\,0.046]}$ &
$0.722_{\,[0.718,\,0.725]}$ &
$0.503_{\,[0.499,\,0.507]}$ &
$0.498_{\,[0.491,\,0.505]}$ \\
\textsc{Gemini-Weak} & 2645 &
$0.379_{\,[0.378,\,0.380]}$ &
$0.720_{\,[0.717,\,0.723]}$ &
$0.179_{\,[0.178,\,0.181]}$ &
$0.038_{\,[0.036,\,0.040]}$ &
$0.648_{\,[0.644,\,0.652]}$ &
$0.502_{\,[0.501,\,0.504]}$ &
$0.501_{\,[0.498,\,0.504]}$ \\
\bottomrule
\end{tabular}%
}
\end{table*}

\subsection{Metric-level tradeoffs and Retrieval baseline behaviour }
\label{sec:metric_tradeoffs}

Table~\ref{tab:main_results} shows that systems differ not only in overall \cbs{}, but also in their metric profiles. \textsc{Gemini-Direct}, \textsc{Gemini-Retrieval}, and \textsc{AI-Researcher} all obtain high \rcf{} scores (0.746, 0.745, and 0.740), indicating coherent and well-formed reasoning traces. However, their \msi{} scores remain very low (0.038, 0.040, and 0.039), suggesting that these outputs are often better at organizing reasoning than at specifying a concrete scientific mechanism. \textsc{ChemDFM-8B} shows a different profile. It achieves the best generated scores on \hpa{} (0.257), \msi{} (0.171), and \pdq{} (0.562), but has a lower \rcf{} score (0.554). This suggests that the chemistry-specialized model is better at proposing structured, domain-grounded hypotheses, even if its reasoning traces are less specific. The reference set remains strongest overall because it balances these dimensions more effectively than any generated system. These results reveal a separation between reasoning fluency and scientific specificity. Strong general-purpose reasoning behavior does not necessarily imply strong scientific problem solving. Co-scientists also need domain grounding in failure modes, meaningful interventions, and plausible mechanisms. This supports the value of adapting strong reasoners to scientific corpora where mechanistic detail and problem decomposition are central. {Gemini-Retrieval} is nearly tied with \textsc{Gemini-Direct} in overall \cbs{} but has a slightly higher \msi{} score (0.040 vs.\ 0.038). This suggests that retrieval adds some mechanistic context, but generic retrieval alone does not close the gap to domain-specialized models or the literature-derived reference.


\subsection{Error taxonomy}
\label{sec:error_taxonomy}

To make the benchmark more interpretable, we group outputs into recurring failure modes using threshold rules over the six metric dimensions. Each entry in Table~\ref{tab:error_taxonomy} is the fraction of outputs from that system assigned to the corresponding failure mode. For example, high alignment, weak mechanism captures outputs with relatively strong \hpa{} but insufficient \msi{}; verbose, weak decomposition captures outputs with strong \rcf{} but weak \pdq{}; fluent but misaligned captures outputs with high \rcf{} and low \hpa{}; and specific but not novel captures outputs with high \msi{} and low \sns{}. Table~\ref{tab:error_taxonomy} shows that verbose, weak decomposition and fluent but misaligned outputs are the most common failure modes overall. \textsc{ChemDFM-8B} has the highest rate of high alignment, weak mechanism outputs (0.220), which means that it often addresses the problem but still lacks enough mechanistic detail in some cases. This does not contradict Table~\ref{tab:main_results}, where \textsc{ChemDFM-8B} has the highest generated \msi{} score; rather, it shows that even the strongest generated system still has a mechanism gap relative to literature-derived hypotheses. By contrast, \textsc{Open Co-Scientist}, \textsc{AI-Researcher}, and \textsc{Gemini-Direct} show high rates of verbose, weak decomposition and fluent but misaligned outputs, suggesting that their reasoning traces can be coherent without being sharply tied to the problem structure.

\begin{table*}[t]
\centering
\scriptsize
\setlength{\tabcolsep}{4pt}
\renewcommand{\arraystretch}{1.08}
\resizebox{\textwidth}{!}{%
\begin{tabular}{lccccccc}
\toprule
\textbf{Error category} & \textbf{Overall} &
\textbf{Ref.} & \textbf{ChemDFM-8B} &
\textbf{G-Direct} & \textbf{G-Weak} &
\textbf{AI-Res.} & \textbf{Open-CS} \\
\midrule
High alignment, weak mechanism
  & 0.067 & 0.015 & 0.220 & 0.084 & 0.037 & 0.062 & 0.004 \\
Novel but implausible
  & 0.014 & 0.000 & 0.272 & 0.000 & 0.000 & 0.000 & 0.000 \\
Verbose, weak decomposition
  & 0.178 & 0.000 & 0.002 & 0.203 & 0.179 & 0.218 & 0.224 \\
Fluent but misaligned
  & 0.175 & 0.000 & \textit{0.010}
  & 0.226 & 0.201 & 0.220 & 0.228 \\
Specific but not novel
  & 0.024 & 0.041 & 0.082
  & 0.008 & 0.011 & 0.009 & 0.009 \\
\bottomrule
\end{tabular}%
}
\caption{
Error taxonomy on \bench{}. Two failure modes dominate: high alignment, weak mechanism
(concentrated in \textsc{ChemDFM-8B}) and verbose, weak decomposition (common in Gemini and agentic systems). Fluent but misaligned (\rcf{}$>$0.65 and \hpa{}$<$0.15) and specific but not novel (\msi{}$>$0.15 and \sns{}$<$0.70).
}
\label{tab:error_taxonomy}
\end{table*}

\begin{figure}[htbp]
    \centering
    \begin{minipage}[c]{0.60\linewidth}
        \centering
        \includegraphics[width=\linewidth]{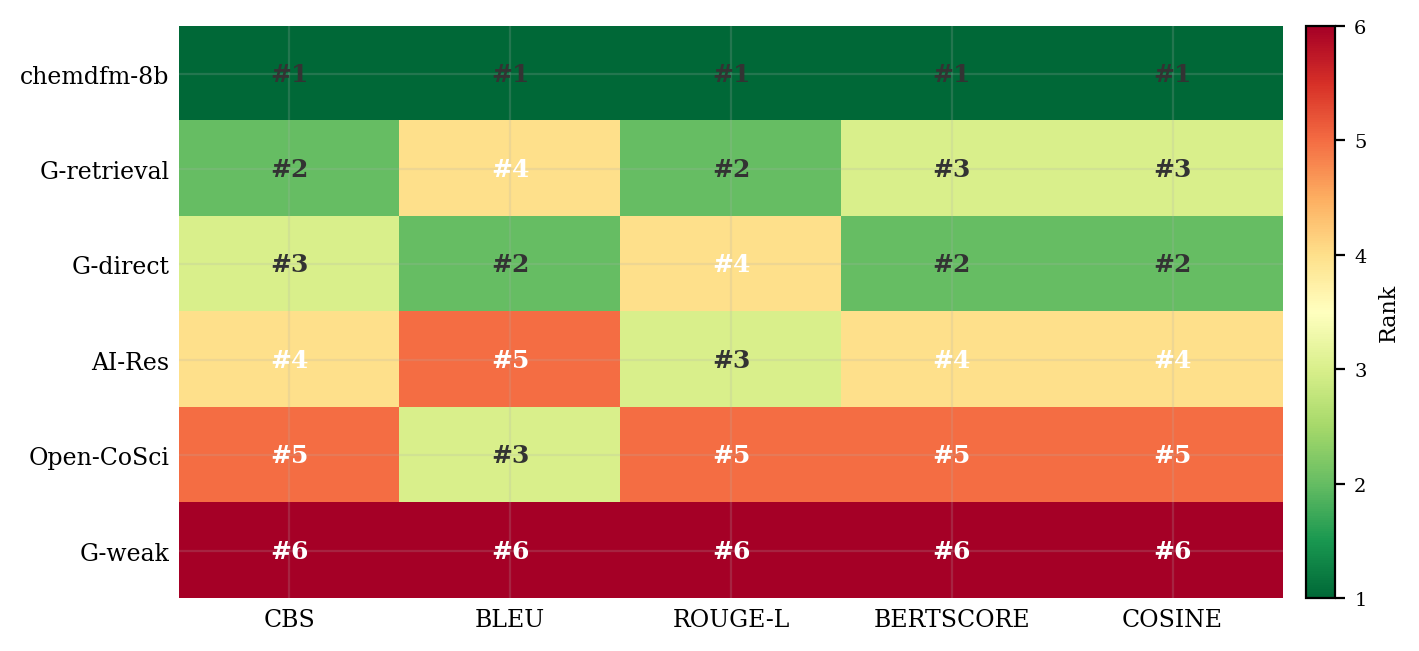}
    \end{minipage}
    \hspace{0.01\linewidth}
    \begin{minipage}[c]{0.36\linewidth}
        \caption{
        Rank heatmap comparing \cbs{} with generic similarity metrics. Lower rank is better. Generic metrics recover part of the benchmark ordering, but not all domain-grounded distinctions: \textsc{ChemDFM-8B} remains first, while local reversals persist, especially between \textsc{Gemini-Retrieval} and \textsc{Gemini-Direct}.
        }
        \label{fig:2}
    \end{minipage}
\end{figure}

\subsection{Validation against reference-based metrics}
\label{sec:generic_validation}

It is possible that \cbs{} is reproducing what standard text-similarity metrics already show (Figure~\ref{fig:2}). To test this, we compare the system-level rankings under \cbs{} with reference-based text-similarity metrics: BLEU
\citep{2018arXiv180408771P,2019arXiv190604903T}, ROUGE-L \citep{shi-etal-2025-judging,2018arXiv180301937G}, BERTScore \citep{2019arXiv190409675Z}, and embedding cosine similarity \citep{2024arXiv240305440S}. Each generated hypothesis is compared against the literature-derived solution hypothesis extracted from the corresponding source paper, using a 500-example subset with 2,149 matched generated rows. Table~\ref{tab:generic_metric_compare} shows that generic metrics partially recover the benchmark ranking, but do not fully reproduce it. \textsc{ChemDFM-8B} remains strongest under all metrics, indicating that the best domain-specialized system is also relatively close to the literature hypotheses. However, agreement is weaker for lexical overlap metrics than for semantic similarity metrics. In particular, Kendall's $\tau$ \citep{2015arXiv150701427S} between \cbs{} and BLEU is 0.60, while agreement with ROUGE-L, BERTScore, and cosine similarity is higher at 0.867. This suggests that these metrics capture semantic closeness to literature hypotheses, but not the full evaluative signal. We also observe that standard metrics do not fully recover the benchmark's local distinctions. \textsc{Gemini-Retrieval} ranks above \textsc{Gemini-Direct} under \cbs{}, but the ordering changes under some similarity-based metrics. This means that conventional metrics capture semantic closeness to literature hypotheses, but do not fully capture the domain-grounded distinctions that our benchmark is designed to measure, such as mechanistic depth and intervention plausibility.


\begin{table*}[t]
\centering

\begin{minipage}[t]{0.48\textwidth}
\centering
\small
\setlength{\tabcolsep}{5pt}
\renewcommand{\arraystretch}{1.06}
\begin{tabular}{lcc}
\toprule
\textbf{Metric} & \textbf{Kendall $\tau$ vs CBS} & \textbf{$p$-value} \\
\midrule
BLEU        & 0.600 & 0.1361 \\
ROUGE-L     & 0.867 & 0.0167 \\
BERTScore   & 0.867 & 0.0167 \\
Cosine sim. & 0.867 & 0.0167 \\
\bottomrule
\end{tabular}
\captionof{table}{
Agreement between \cbs{} and conventional metrics. Conventional metrics partially recover the benchmark ranking, but the weaker agreement for BLEU and the remaining local ranking reversals suggest that \bench{} captures distinctions beyond surface overlap. 
}
\label{tab:generic_metric_compare}
\end{minipage}
\hfill
\begin{minipage}[t]{0.48\textwidth}
\centering
\small
\setlength{\tabcolsep}{4pt}
\renewcommand{\arraystretch}{1.06}
\begin{tabular}{lc}
\toprule
\textbf{Statistic} & \textbf{Value} \\
\midrule
CBS-Gemini 2.5 Pro agreement   & 0.40 \\
CBS-Gemini Flash agreement & 0.60 \\
Pro-Flash agreement        & 0.32 \\
Mean order-flip rate        & 0.236 \\
\bottomrule
\end{tabular}
\captionof{table}{
Summary of pairwise LLM-judge validation. Gemini Flash agrees with \cbs{} more often than Gemini 2.5 Pro, while inter-judge agreement remains limited.}
\label{tab:judge_summary}
\end{minipage}

\end{table*}

\subsection{Pairwise LLM-judge validation}
\label{sec:judge_validation}

We also performed a validation experiment using pairwise LLM judging. We use Gemini 2.5 Pro and Gemini 2.5 Flash. For each system pair, we compare outputs on 50 common problems using two prompt orders (A/B and B/A) to reduce order bias \citep{2024arXiv241002736Y,2025arXiv250622316L,shi-etal-2025-judging}. The full judge prompt is provided in Appendix~\ref{app:judge_prompt}. Each comparison is run twice with swapped order to reduce position bias. Across the tested pairs, agreement between \cbs{} and Gemini Flash is 0.60, while agreement between \cbs{} and Gemini 2.5 Pro is 0.40. Agreement between the two judges themselves is 0.32, and the mean order-flip rate is 0.236. These numbers indicate that pairwise LLM judgment is informative in this setting, but also noisy and sensitive to both judge choice and presentation order. Gemini Flash is more decisive and more often agrees with the benchmark ranking. Gemini 2.5 Pro, by contrast, is highly tie-prone, which lowers its raw agreement with \cbs{} even when it rarely makes strong contradictory judgments. For example, in the \textsc{ChemDFM-8B} versus \textsc{Gemini-Direct} comparison, Gemini 2.5 Pro returns 36 ties out of 50 cases, while Gemini Flash returns only 5 ties and strongly prefers \textsc{ChemDFM-8B}. Similar behavior appears in the \textsc{ChemDFM-8B} versus \textsc{AI-Researcher} comparison. The results are consistent with the main benchmark conclusions, especially for clearer system gaps, but they also reinforce an important methodological point: external LLM judgment in scientific settings is itself unstable and should not be treated as ground truth.

\subsection{Metric-gaming test}
\label{sec:gaming_validation}

We also tested whether the benchmark can be manipulated by adversarial strategies (Table~\ref{tab:gaming_summary}). We generate 75 adversarial examples for each of three attack styles: jargon stuffing \citep{2024arXiv240806509Y}, problem mirroring \citep{2024arXiv240214963Y}, and verbose fake reasoning \citep{2025arXiv250722940J}. These attacks were designed to produce outputs that look aligned to the problem while remaining weak in actual scientific content. We observed that none of these attacks improves the overall \cbs{} score. All adversarial styles reduce \cbs{} relative to the original outputs: jargon stuffing by 0.0348, problem mirroring by 0.0353, and verbose fake reasoning by 0.0543. However, some local dimensions can be manipulated. \hpa{} increases under all three attack styles, with the largest increase coming from problem mirroring (+0.1327). This is expected since mirroring key problem terms can make an output appear more aligned to the input problem. However, these same attacks are strongly penalized on \msi{} and \ip{}. Jargon stuffing reduces \msi{} by 0.1307 and \ip{} by 0.1580, while verbose fake reasoning reduces \msi{} by 0.1486 and \ip{} by 0.1310. This means that although individual dimensions can be partially gamed, the full benchmark is substantially more resistant than any single component alone.

\begin{table}[t]
\centering
\small
\setlength{\tabcolsep}{4pt}
\renewcommand{\arraystretch}{1.06}
\begin{tabular}{lcccc}
\toprule
\textbf{Type of Attack} & $\Delta$HPA & $\Delta$MSI & $\Delta$IP & $\Delta$CBS \\
\midrule
Jargon stuffing       & +0.0679 & -0.1307 & -0.1580 & -0.0348 \\
Problem mirroring     & +0.1327 & -0.1295 & -0.0886 & -0.0353 \\
Verbose fake reasoning& +0.0602 & -0.1486 & -0.1310 & -0.0543 \\
\bottomrule
\end{tabular}
\caption{
Metric-gaming stress test. Simple adversarial attacks can inflate local alignment signals, especially \hpa{}, but they reduce mechanistic specificity, plausibility, and overall \cbs{}. This suggests that the aggregate benchmark is more stable than any individual dimension alone.
}
\label{tab:gaming_summary}
\end{table}

\section{Discussion and Limitations}
\label{sec:discussion_limitations}
Our results show that \bench{} is non-trivial: literature-derived hypotheses remain stronger overall than current AI co-scientist outputs. The benchmark separates systems in an interpretable way, especially along the axis of reasoning fluency versus scientific specificity. Conventional metrics recover only part of the evaluative signal. The benchmark is not trivially broken by simple adversarial strategies, although some individual dimensions are easier to manipulate than others. At the same time, \bench{} evaluates one specific part of scientific reasoning, which is moving from a structured materials problem to a plausible solution hypothesis. Additional ablations on metric redundancy and rare-term contamination are provided in (section ~\ref{app:ablation_results})  The benchmark is constructed from published papers, and the structured fields are automatically extracted, so some examples may contain noise or incomplete structure. Pairwise LLM judging provides useful proxy evidence, but judges show limited agreement and remain sensitive to prompt order. Moreover, the benchmark focuses on hypotheses and mechanisms, rather than explicit tool use or experiment planning. In real material workflows, a useful output may also need to specify next-step analyses such as DFT calculations, electrochemical protocols, or characterization methods. Incorporating these decisions is an important direction for future work. Discussion of metric validity, evidence-strength limitations in \ip{}, retrieval/direct parity, and reference \hpa{} behavior is provided in Appendix~\ref{app:metric_validity}.

\section{Ablation Study}
\label{app:ablation_results}

We performed two additional checks. First, the six metrics are related but not redundant: pairwise correlations are generally small, with near-zero correlation between \rcf{} and \sns{} ($\rho=0.0045$) and between \hpa{} and \sns{} ($\rho=0.0229$), while \hpa{} and \msi{} ($\rho=0.2191$) and \msi{} and \pdq{} ($\rho=0.2211$) show only modest positive correlation. This suggests that the metrics belong to the same task but capture different aspects of scientific reasoning, so strong fluent reasoning does not automatically imply strong mechanism or alignment. Second, since the benchmark is built from published papers, contamination and memorization can occur. We mitigate this by using reference-free metrics that score outputs against the input problem and internal structure rather than by direct answer matching. Rare-term analysis suggests that highly specific mechanistic detail is uncommon in the corpus: the mean rare-term count is 0.021, 98\% of examples contain zero such terms, and only 0.1\% contain two or more. This reduces the risk that benchmark performance is dominated by obvious memorization.

\subsection{CBS Weight Sensitivity}
\label{app:weight_ablation}
 
Table~\ref{tab:weight_ablation} reports system rankings under four alternative \cbs{} weight schemes. The top two systems, \textsc{Reference} and \textsc{ChemDFM-8B}, are stable across all schemes, and the bottom system, \textsc{Gemini-Weak}, is also stable. The only instability occurs between \textsc{Gemini-Direct} and \textsc{Gemini-Retrieval} under the \msi{}-heavy scheme. This is interpretable: when mechanistic specificity is weighted at 0.40, the small \msi{} advantage of \textsc{Gemini-Retrieval} (0.040 vs.\ 0.038) becomes decisive. Overall, the rankings are stable under reasonable weight perturbations, with Kendall's $\tau \geq 0.905$ relative to the default ranking.
 
\begin{table}[ht]
\centering
\small
\caption{
    \cbs{} scores under alternative weight schemes.
    Default: $w = [0.20, 0.20, 0.18, 0.15, 0.15, 0.12]$
    for (\rcf{}, \hpa{}, \msi{}, \sns{}, \ip{}, \pdq{}).
    Rankings are stable except for the \textsc{Gemini-Direct}/
    \textsc{Gemini-Retrieval} swap under \msi{}-heavy weighting.
}
\label{tab:weight_ablation}
\setlength{\tabcolsep}{5pt}
\begin{tabular}{lcccc}
\toprule
\textbf{System} &
\textbf{Default} & \textbf{Uniform} &
\textbf{\rcf{}-heavy} & \textbf{\msi{}-heavy} \\
\midrule
\textsc{Reference}        & 0.467 & 0.482 & 0.561 & 0.405 \\
\textsc{ChemDFM-8B}       & 0.427 & 0.442 & 0.473 & 0.366 \\
\textsc{Gemini-Direct}    & 0.388 & 0.330 & 0.446 & 0.248 \\
\textsc{Gemini-Retrieval} & 0.388 & 0.330 & 0.446 & \textbf{0.249} \\
\textsc{AI-Researcher}    & 0.386 & 0.328 & 0.444 & 0.247 \\
\textsc{Open Co-Scientist}& 0.379 & 0.323 & 0.440 & 0.245 \\
\textsc{Gemini-Weak}      & 0.379 & 0.323 & 0.434 & 0.243 \\
\midrule
\multicolumn{5}{l}{Rank correlation with default ranking (Kendall's $\tau$):} \\
vs.\ Uniform     & --- & 0.905 & --- & --- \\
vs.\ \rcf{}-heavy   & --- & --- & 1.000 & --- \\
vs.\ \msi{}-heavy   & --- & --- & --- & 0.905 \\
\bottomrule
\end{tabular}
\end{table}

\subsection{Additional Validations}
\label{app:additional_validity_checks}

We ran three additional checks to clarify potential weaknesses of the benchmark. The reliability of the \ip{} evidence-strength field, the quality of automatically extracted instances, and the near-parity between \textsc{Gemini-Direct} and \textsc{Gemini-Retrieval}.

\paragraph{Evidence-strength field in \ip{}.}

The \ip{} metric includes an evidence-strength subscore extracted from the source paper. Across the 2{,}645 benchmark instances, 61.3\% are labeled strong, 38.4\% are labeled moderate, 0.2\% are labeled unknown, and 0.1\% are labeled weak. These map to evidence subscores of 1.00, 0.65, 0.35, and 0.25, respectively. The mean evidence subscore is 0.864, with only 0.2\% of examples receiving the unknown fallback. This suggests that the \ip{} evidence component is usually supported by an explicit evidence label rather than dominated by missing-field defaults.

\paragraph{Instance-quality spot check.}

We also performed a spot check of 50 randomly sampled instances using GPT-5.5 as the evaluator. We found 31 responses were validly parsed. The pass rates were 96.8\% for problem-statement correctness, 100.0\% for failure-mode specificity, 96.8\% for intervention relevance, and 87.1\% for mechanism coherence. Overall, 83.9\% of valid instances passed all four criteria, and 87.1\% were rated high quality. Because the evaluator is an LLM rather than a human expert, we treat this as a consistency check rather than independent human validation.

\paragraph{Retrieval-direct parity.}
\textsc{Gemini-Retrieval} and \textsc{Gemini-Direct} achieve nearly identical \cbs{} scores (0.3877 vs.\ 0.3878). Retrieval slightly increases \msi{} (+0.0012) and \ip{} (+0.0002), but slightly decreases \rcf{} (-0.0006) and \hpa{} (-0.0011). This suggests that retrieval adds a small amount of mechanistic context, but does not substantially change the quality of the generated hypotheses. In this benchmark, generic retrieval alone is therefore not sufficient to close the gap to domain-specialized systems such as \textsc{ChemDFM-8B}.

\begin{table}[h]
\centering
\small
\caption{Instance-quality spot check on 31 valid randomly sampled instances.}
\label{tab:spotcheck}
\begin{tabular}{lc}
\toprule
\textbf{Criterion} & \textbf{Pass rate} \\
\midrule
Problem statement correct & 96.8\% \\
Failure mode specific & 100.0\% \\
Intervention relevant & 96.8\% \\
Mechanism coherent & 87.1\% \\
All four correct & 83.9\% \\
Overall high quality & 87.1\% \\
\bottomrule
\end{tabular}
\end{table}

\section{Acknowledgments}

This work was supported by the U.S. Department of Energy, Office of Science, Office of Advanced Scientific Computing Research and Office of Basic Energy Sciences, Scientific Discovery through Advanced Computing (SciDAC) program under the FORUM-AI project. Work performed at the Center for Nanoscale Materials, a U.S. Department of Energy Office of Science User Facility, was supported by the U.S. DOE, Office of Basic Energy Sciences, under Contract No. DE-AC02-06CH11357.

\bibliographystyle{plainnat}
\bibliography{references}

\begin{thebibliography}{31}
\providecommand{\natexlab}[1]{#1}
\providecommand{\url}[1]{\texttt{#1}}
\expandafter\ifx\csname urlstyle\endcsname\relax
  \providecommand{\doi}[1]{doi: #1}\else
  \providecommand{\doi}{doi: \begingroup \urlstyle{rm}\Url}\fi

\bibitem[{Alampara} et~al.(2024){Alampara}, {Schilling-Wilhelmi}, {R{\'\i}os-Garc{\'\i}a}, {Mandal}, {Khetarpal}, {Singh Grover}, {Anoop Krishnan}, and {Maik Jablonka}]{2024arXiv241116955A}
Nawaf {Alampara}, Mara {Schilling-Wilhelmi}, Marti{\~n}o {R{\'\i}os-Garc{\'\i}a}, Indrajeet {Mandal}, Pranav {Khetarpal}, Hargun {Singh Grover}, N.~M. {Anoop Krishnan}, and Kevin {Maik Jablonka}.
\newblock {Probing the limitations of multimodal language models for chemistry and materials research}.
\newblock \emph{arXiv e-prints}, art. arXiv:2411.16955, November 2024.
\newblock \doi{10.48550/arXiv.2411.16955}.

\bibitem[{Chehbouni} et~al.(2025){Chehbouni}, {Haddou}, {Kit Cheung}, and {Farnadi}]{2025arXiv250818076C}
Khaoula {Chehbouni}, Mohammed {Haddou}, Jackie~Chi {Kit Cheung}, and Golnoosh {Farnadi}.
\newblock {Neither Valid nor Reliable? Investigating the Use of LLMs as Judges}.
\newblock \emph{arXiv e-prints}, art. arXiv:2508.18076, August 2025.
\newblock \doi{10.48550/arXiv.2508.18076}.

\bibitem[{Chen} et~al.(2024){Chen}, {Chen}, {Ning}, {Zhang}, {Wang}, {Yu}, {Li}, {Liao}, {Wei}, {Lu}, {Dey}, {Xue}, {Baker}, {Burns}, {Adu-Ampratwum}, {Huang}, {Ning}, {Gao}, {Su}, and {Sun}]{2024arXiv241005080C}
Ziru {Chen}, Shijie {Chen}, Yuting {Ning}, Qianheng {Zhang}, Boshi {Wang}, Botao {Yu}, Yifei {Li}, Zeyi {Liao}, Chen {Wei}, Zitong {Lu}, Vishal {Dey}, Mingyi {Xue}, Frazier~N. {Baker}, Benjamin {Burns}, Daniel {Adu-Ampratwum}, Xuhui {Huang}, Xia {Ning}, Song {Gao}, Yu~{Su}, and Huan {Sun}.
\newblock {ScienceAgentBench: Toward Rigorous Assessment of Language Agents for Data-Driven Scientific Discovery}.
\newblock \emph{arXiv e-prints}, art. arXiv:2410.05080, October 2024.
\newblock \doi{10.48550/arXiv.2410.05080}.

\bibitem[{Dehghani} et~al.(2021){Dehghani}, {Tay}, {Gritsenko}, {Zhao}, {Houlsby}, {Diaz}, {Metzler}, and {Vinyals}]{2021arXiv210707002D}
Mostafa {Dehghani}, Yi~{Tay}, Alexey~A. {Gritsenko}, Zhe {Zhao}, Neil {Houlsby}, Fernando {Diaz}, Donald {Metzler}, and Oriol {Vinyals}.
\newblock {The Benchmark Lottery}.
\newblock \emph{arXiv e-prints}, art. arXiv:2107.07002, July 2021.
\newblock \doi{10.48550/arXiv.2107.07002}.

\bibitem[{Ganesan}(2018)]{2018arXiv180301937G}
Kavita {Ganesan}.
\newblock {ROUGE 2.0: Updated and Improved Measures for Evaluation of Summarization Tasks}.
\newblock \emph{arXiv e-prints}, art. arXiv:1803.01937, March 2018.
\newblock \doi{10.48550/arXiv.1803.01937}.

\bibitem[{Gottweis} et~al.(2025){Gottweis}, {Weng}, {Daryin}, {Tu}, {Palepu}, {Sirkovic}, {Myaskovsky}, {Weissenberger}, {Rong}, {Tanno}, {Saab}, {Popovici}, {Blum}, {Zhang}, {Chou}, {Hassidim}, {Gokturk}, {Vahdat}, {Kohli}, {Matias}, {Carroll}, {Kulkarni}, {Tomasev}, {Guan}, {Dhillon}, {Dhaval Vaishnav}, {Lee}, {Costa}, {Penad{\'e}s}, {Peltz}, {Xu}, {Pawlosky}, {Karthikesalingam}, and {Natarajan}]{2025arXiv250218864G}
Juraj {Gottweis}, Wei-Hung {Weng}, Alexander {Daryin}, Tao {Tu}, Anil {Palepu}, Petar {Sirkovic}, Artiom {Myaskovsky}, Felix {Weissenberger}, Keran {Rong}, Ryutaro {Tanno}, Khaled {Saab}, Dan {Popovici}, Jacob {Blum}, Fan {Zhang}, Katherine {Chou}, Avinatan {Hassidim}, Burak {Gokturk}, Amin {Vahdat}, Pushmeet {Kohli}, Yossi {Matias}, Andrew {Carroll}, Kavita {Kulkarni}, Nenad {Tomasev}, Yuan {Guan}, Vikram {Dhillon}, Eeshit {Dhaval Vaishnav}, Byron {Lee}, Tiago R~D {Costa}, Jos{\'e}~R {Penad{\'e}s}, Gary {Peltz}, Yunhan {Xu}, Annalisa {Pawlosky}, Alan {Karthikesalingam}, and Vivek {Natarajan}.
\newblock {Towards an AI co-scientist}.
\newblock \emph{arXiv e-prints}, art. arXiv:2502.18864, February 2025.
\newblock \doi{10.48550/arXiv.2502.18864}.

\bibitem[{Gu} et~al.(2024){Gu}, {Jiang}, {Shi}, {Tan}, {Zhai}, {Xu}, {Li}, {Shen}, {Ma}, {Liu}, {Wang}, {Zhang}, {Wang}, {Gao}, {Ni}, and {Guo}]{2024arXiv241115594G}
Jiawei {Gu}, Xuhui {Jiang}, Zhichao {Shi}, Hexiang {Tan}, Xuehao {Zhai}, Chengjin {Xu}, Wei {Li}, Yinghan {Shen}, Shengjie {Ma}, Honghao {Liu}, Saizhuo {Wang}, Kun {Zhang}, Yuanzhuo {Wang}, Wen {Gao}, Lionel {Ni}, and Jian {Guo}.
\newblock {A Survey on LLM-as-a-Judge}.
\newblock \emph{arXiv e-prints}, art. arXiv:2411.15594, November 2024.
\newblock \doi{10.48550/arXiv.2411.15594}.

\bibitem[{Jiao} et~al.(2025){Jiao}, {Zhang}, and {Li}]{2025arXiv250722940J}
Rui {Jiao}, Yue {Zhang}, and Jinku {Li}.
\newblock {Trustworthy Reasoning: Evaluating and Enhancing Factual Accuracy in LLM Intermediate Thought Processes}.
\newblock \emph{arXiv e-prints}, art. arXiv:2507.22940, July 2025.
\newblock \doi{10.48550/arXiv.2507.22940}.

\bibitem[{Kumbhar} et~al.(2025){Kumbhar}, {Mishra}, {Coutinho}, {Handa}, {Iquebal}, and {Baral}]{2025arXiv250113299K}
Shrinidhi {Kumbhar}, Venkatesh {Mishra}, Kevin {Coutinho}, Divij {Handa}, Ashif {Iquebal}, and Chitta {Baral}.
\newblock {Hypothesis Generation for Materials Discovery and Design Using Goal-Driven and Constraint-Guided LLM Agents}.
\newblock \emph{arXiv e-prints}, art. arXiv:2501.13299, January 2025.
\newblock \doi{10.48550/arXiv.2501.13299}.

\bibitem[{Li} et~al.(2025){Li}, {Dou}, {Shao}, {Chen}, and {Hu}]{2025arXiv250622316L}
Qingquan {Li}, Shaoyu {Dou}, Kailai {Shao}, Chao {Chen}, and Haixiang {Hu}.
\newblock {Evaluating Scoring Bias in LLM-as-a-Judge}.
\newblock \emph{arXiv e-prints}, art. arXiv:2506.22316, June 2025.
\newblock \doi{10.48550/arXiv.2506.22316}.

\bibitem[{Liu} et~al.(2025{\natexlab{a}}){Liu}, {Hu}, {Ye}, {Xu}, {Srolovitz}, and {Wen}]{2025arXiv250510852L}
Siyu {Liu}, Bo~{Hu}, Beilin {Ye}, Jiamin {Xu}, David~J. {Srolovitz}, and Tongqi {Wen}.
\newblock {MatTools: Benchmarking Large Language Models for Materials Science Tools}.
\newblock \emph{arXiv e-prints}, art. arXiv:2505.10852, May 2025{\natexlab{a}}.
\newblock \doi{10.48550/arXiv.2505.10852}.

\bibitem[{Liu} et~al.(2025{\natexlab{b}}){Liu}, {Yang}, {Xie}, {Ni}, {Gao}, {Li}, {Tang}, {Ouyang}, {Cambria}, and {Zhou}]{2025arXiv250321248L}
Yujie {Liu}, Zonglin {Yang}, Tong {Xie}, Jinjie {Ni}, Ben {Gao}, Yuqiang {Li}, Shixiang {Tang}, Wanli {Ouyang}, Erik {Cambria}, and Dongzhan {Zhou}.
\newblock {ResearchBench: Benchmarking LLMs in Scientific Discovery via Inspiration-Based Task Decomposition}.
\newblock \emph{arXiv e-prints}, art. arXiv:2503.21248, March 2025{\natexlab{b}}.
\newblock \doi{10.48550/arXiv.2503.21248}.

\bibitem[{Longjohn} et~al.(2024){Longjohn}, {Kelly}, {Singh}, and {Smyth}]{2024arXiv241024100L}
Rachel {Longjohn}, Markelle {Kelly}, Sameer {Singh}, and Padhraic {Smyth}.
\newblock {Benchmark Data Repositories for Better Benchmarking}.
\newblock \emph{arXiv e-prints}, art. arXiv:2410.24100, October 2024.
\newblock \doi{10.48550/arXiv.2410.24100}.

\bibitem[{Mirza} et~al.(2024){Mirza}, {Alampara}, {Kunchapu}, {R{\'\i}os-Garc{\'\i}a}, {Emoekabu}, {Krishnan}, {Gupta}, {Schilling-Wilhelmi}, {Okereke}, {Aneesh}, {Elahi}, {Asgari}, {Eberhardt}, {Elbeheiry}, {Gil}, {Greiner}, {Holick}, {Glaubitz}, {Hoffmann}, {Ibrahim}, {Klepsch}, {K{\"o}ster}, {Kreth}, {Meyer}, {Miret}, {Peschel}, {Ringleb}, {Roesner}, {Schreiber}, {Schubert}, {Stafast}, {Wonanke}, {Pieler}, {Schwaller}, and {Maik Jablonka}]{2024arXiv240401475M}
Adrian {Mirza}, Nawaf {Alampara}, Sreekanth {Kunchapu}, Marti{\~n}o {R{\'\i}os-Garc{\'\i}a}, Benedict {Emoekabu}, Aswanth {Krishnan}, Tanya {Gupta}, Mara {Schilling-Wilhelmi}, Macjonathan {Okereke}, Anagha {Aneesh}, Amir~Mohammad {Elahi}, Mehrdad {Asgari}, Juliane {Eberhardt}, Hani~M. {Elbeheiry}, Mar{\'\i}a~Victoria {Gil}, Maximilian {Greiner}, Caroline~T. {Holick}, Christina {Glaubitz}, Tim {Hoffmann}, Abdelrahman {Ibrahim}, Lea~C. {Klepsch}, Yannik {K{\"o}ster}, Fabian~Alexander {Kreth}, Jakob {Meyer}, Santiago {Miret}, Jan~Matthias {Peschel}, Michael {Ringleb}, Nicole {Roesner}, Johanna {Schreiber}, Ulrich~S. {Schubert}, Leanne~M. {Stafast}, Dinga {Wonanke}, Michael {Pieler}, Philippe {Schwaller}, and Kevin {Maik Jablonka}.
\newblock {Are large language models superhuman chemists?}
\newblock \emph{arXiv e-prints}, art. arXiv:2404.01475, April 2024.
\newblock \doi{10.48550/arXiv.2404.01475}.

\bibitem[{Mitchener} et~al.(2025){Mitchener}, {Laurent}, {Andonian}, {Tenmann}, {Narayanan}, {Wellawatte}, {White}, {Sani}, and {Rodriques}]{2025arXiv250300096M}
Ludovico {Mitchener}, Jon~M {Laurent}, Alex {Andonian}, Benjamin {Tenmann}, Siddharth {Narayanan}, Geemi~P {Wellawatte}, Andrew {White}, Lorenzo {Sani}, and Samuel~G {Rodriques}.
\newblock {BixBench: a Comprehensive Benchmark for LLM-based Agents in Computational Biology}.
\newblock \emph{arXiv e-prints}, art. arXiv:2503.00096, February 2025.
\newblock \doi{10.48550/arXiv.2503.00096}.

\bibitem[{Post}(2018)]{2018arXiv180408771P}
Matt {Post}.
\newblock {A Call for Clarity in Reporting BLEU Scores}.
\newblock \emph{arXiv e-prints}, art. arXiv:1804.08771, April 2018.
\newblock \doi{10.48550/arXiv.1804.08771}.

\bibitem[{Prasad Majumder} et~al.(2024){Prasad Majumder}, {Surana}, {Agarwal}, {Dalvi Mishra}, {Meena}, {Prakhar}, {Vora}, {Khot}, {Sabharwal}, and {Clark}]{2024arXiv240701725P}
Bodhisattwa {Prasad Majumder}, Harshit {Surana}, Dhruv {Agarwal}, Bhavana {Dalvi Mishra}, Abhijeetsingh {Meena}, Aryan {Prakhar}, Tirth {Vora}, Tushar {Khot}, Ashish {Sabharwal}, and Peter {Clark}.
\newblock {DiscoveryBench: Towards Data-Driven Discovery with Large Language Models}.
\newblock \emph{arXiv e-prints}, art. arXiv:2407.01725, July 2024.
\newblock \doi{10.48550/arXiv.2407.01725}.

\bibitem[{Raji} et~al.(2021){Raji}, {Bender}, {Paullada}, {Denton}, and {Hanna}]{2021arXiv211115366R}
Inioluwa~Deborah {Raji}, Emily~M. {Bender}, Amandalynne {Paullada}, Emily {Denton}, and Alex {Hanna}.
\newblock {AI and the Everything in the Whole Wide World Benchmark}.
\newblock \emph{arXiv e-prints}, art. arXiv:2111.15366, November 2021.
\newblock \doi{10.48550/arXiv.2111.15366}.

\bibitem[Shi et~al.(2025)Shi, Ma, Liang, Diao, Ma, and Vosoughi]{shi-etal-2025-judging}
Lin Shi, Chiyu Ma, Wenhua Liang, Xingjian Diao, Weicheng Ma, and Soroush Vosoughi.
\newblock Judging the judges: A systematic study of position bias in {LLM}-as-a-judge.
\newblock In Kentaro Inui, Sakriani Sakti, Haofen Wang, Derek~F. Wong, Pushpak Bhattacharyya, Biplab Banerjee, Asif Ekbal, Tanmoy Chakraborty, and Dhirendra~Pratap Singh, editors, \emph{Proceedings of the 14th International Joint Conference on Natural Language Processing and the 4th Conference of the Asia-Pacific Chapter of the Association for Computational Linguistics}, pages 292--314, Mumbai, India, December 2025. The Asian Federation of Natural Language Processing and The Association for Computational Linguistics.
\newblock ISBN 979-8-89176-298-5.
\newblock \doi{10.18653/v1/2025.ijcnlp-long.18}.
\newblock URL \url{https://aclanthology.org/2025.ijcnlp-long.18/}.

\bibitem[{Steck} et~al.(2024){Steck}, {Ekanadham}, and {Kallus}]{2024arXiv240305440S}
Harald {Steck}, Chaitanya {Ekanadham}, and Nathan {Kallus}.
\newblock {Is Cosine-Similarity of Embeddings Really About Similarity?}
\newblock \emph{arXiv e-prints}, art. arXiv:2403.05440, March 2024.
\newblock \doi{10.48550/arXiv.2403.05440}.

\bibitem[{Stepanov}(2015)]{2015arXiv150701427S}
Alexei {Stepanov}.
\newblock {On the Kendall Correlation Coefficient}.
\newblock \emph{arXiv e-prints}, art. arXiv:1507.01427, July 2015.
\newblock \doi{10.48550/arXiv.1507.01427}.

\bibitem[{Tang} et~al.(2025){Tang}, {Xia}, {Li}, and {Huang}]{2025arXiv250518705T}
Jiabin {Tang}, Lianghao {Xia}, Zhonghang {Li}, and Chao {Huang}.
\newblock {AI-Researcher: Autonomous Scientific Innovation}.
\newblock \emph{arXiv e-prints}, art. arXiv:2505.18705, May 2025.
\newblock \doi{10.48550/arXiv.2505.18705}.

\bibitem[{Tian} et~al.(2024){Tian}, {Gao}, {Dylan Zhang}, {Chen}, {Fan}, {Guo}, {Haas}, {Ji}, {Krongchon}, {Li}, {Liu}, {Luo}, {Ma}, {Tong}, {Trinh}, {Tian}, {Wang}, {Wu}, {Xiong}, {Yin}, {Zhu}, {Lieret}, {Lu}, {Liu}, {Du}, {Tao}, {Press}, {Callan}, {Huerta}, and {Peng}]{2024arXiv240713168T}
Minyang {Tian}, Luyu {Gao}, Shizhuo {Dylan Zhang}, Xinan {Chen}, Cunwei {Fan}, Xuefei {Guo}, Roland {Haas}, Pan {Ji}, Kittithat {Krongchon}, Yao {Li}, Shengyan {Liu}, Di~{Luo}, Yutao {Ma}, Hao {Tong}, Kha {Trinh}, Chenyu {Tian}, Zihan {Wang}, Bohao {Wu}, Yanyu {Xiong}, Shengzhu {Yin}, Minhui {Zhu}, Kilian {Lieret}, Yanxin {Lu}, Genglin {Liu}, Yufeng {Du}, Tianhua {Tao}, Ofir {Press}, Jamie {Callan}, Eliu {Huerta}, and Hao {Peng}.
\newblock {SciCode: A Research Coding Benchmark Curated by Scientists}.
\newblock \emph{arXiv e-prints}, art. arXiv:2407.13168, July 2024.
\newblock \doi{10.48550/arXiv.2407.13168}.

\bibitem[{Tran} et~al.(2019){Tran}, {Tran}, {Nguyen}, {Nguyen}, and {Nguyen}]{2019arXiv190604903T}
Ngoc {Tran}, Hieu {Tran}, Son {Nguyen}, Hoan {Nguyen}, and Tien~N. {Nguyen}.
\newblock {Does BLEU Score Work for Code Migration?}
\newblock \emph{arXiv e-prints}, art. arXiv:1906.04903, June 2019.
\newblock \doi{10.48550/arXiv.1906.04903}.

\bibitem[{Wagstaff}(2012)]{2012arXiv1206.4656W}
Kiri {Wagstaff}.
\newblock {Machine Learning that Matters}.
\newblock \emph{arXiv e-prints}, art. arXiv:1206.4656, June 2012.
\newblock \doi{10.48550/arXiv.1206.4656}.

\bibitem[{Yan} et~al.(2024){Yan}, {Zhu}, {Wang}, {Gui}, and {He}]{2024arXiv240214963Y}
Hanqi {Yan}, Qinglin {Zhu}, Xinyu {Wang}, Lin {Gui}, and Yulan {He}.
\newblock {Mirror: A Multiple-perspective Self-Reflection Method for Knowledge-rich Reasoning}.
\newblock \emph{arXiv e-prints}, art. arXiv:2402.14963, February 2024.
\newblock \doi{10.48550/arXiv.2402.14963}.

\bibitem[{Ye} et~al.(2024){Ye}, {Wang}, {Huang}, {Chen}, {Zhang}, {Moniz}, {Gao}, {Geyer}, {Huang}, {Chen}, {Chawla}, and {Zhang}]{2024arXiv241002736Y}
Jiayi {Ye}, Yanbo {Wang}, Yue {Huang}, Dongping {Chen}, Qihui {Zhang}, Nuno {Moniz}, Tian {Gao}, Werner {Geyer}, Chao {Huang}, Pin-Yu {Chen}, Nitesh~V {Chawla}, and Xiangliang {Zhang}.
\newblock {Justice or Prejudice? Quantifying Biases in LLM-as-a-Judge}.
\newblock \emph{arXiv e-prints}, art. arXiv:2410.02736, October 2024.
\newblock \doi{10.48550/arXiv.2410.02736}.

\bibitem[{Yuan} and {Dasgupta}(2024)]{2024arXiv240806509Y}
Jun {Yuan} and Aritra {Dasgupta}.
\newblock {Fooling SHAP with Output Shuffling Attacks}.
\newblock \emph{arXiv e-prints}, art. arXiv:2408.06509, August 2024.
\newblock \doi{10.48550/arXiv.2408.06509}.

\bibitem[{Zhang} et~al.(2019){Zhang}, {Kishore}, {Wu}, {Weinberger}, and {Artzi}]{2019arXiv190409675Z}
Tianyi {Zhang}, Varsha {Kishore}, Felix {Wu}, Kilian~Q. {Weinberger}, and Yoav {Artzi}.
\newblock {BERTScore: Evaluating Text Generation with BERT}.
\newblock \emph{arXiv e-prints}, art. arXiv:1904.09675, April 2019.
\newblock \doi{10.48550/arXiv.1904.09675}.

\bibitem[{Zhao} et~al.(2025){Zhao}, {Ma}, {Chen}, {Sun}, {Li}, {Xia}, {Chen}, {Xu}, {Zhu}, {Zhu}, {Fan}, {Shen}, {Yu}, and {Chen}]{2025CRPS....602523Z}
Zihan {Zhao}, Da~{Ma}, Lu~{Chen}, Liangtai {Sun}, Zihao {Li}, Yi~{Xia}, Bo~{Chen}, Hongshen {Xu}, Zichen {Zhu}, Su~{Zhu}, Shuai {Fan}, Guodong {Shen}, Kai {Yu}, and Xin {Chen}.
\newblock {Developing ChemDFM as a large language foundation model for chemistry}.
\newblock \emph{Cell Reports Physical Science}, 6\penalty0 (4):\penalty0 102523, April 2025.
\newblock \doi{10.1016/j.xcrp.2025.102523}.

\bibitem[{Zou} et~al.(2023){Zou}, {Liu}, {Miao}, {Fung}, {Chang}, and {Li}]{2023arXiv231008677Z}
Deyu {Zou}, Shikun {Liu}, Siqi {Miao}, Victor {Fung}, Shiyu {Chang}, and Pan {Li}.
\newblock {GeSS: Benchmarking Geometric Deep Learning under Scientific Applications with Distribution Shifts}.
\newblock \emph{arXiv e-prints}, art. arXiv:2310.08677, October 2023.
\newblock \doi{10.48550/arXiv.2310.08677}.

\end{thebibliography}

\appendix
\section{Metrics}
\label{app:metrics}

\subsection{Reasoning Chain Fidelity (\rcf{})}
\label{app:rcf}
 
Steps $\{s_1,\ldots,s_n\}$ are extracted by parsing \texttt{[Begin Step $k$]\ldots[End Step $k$]} tags from
the \texttt{reasoning\_process} field.
 
\paragraph{Progression.}
For each adjacent pair $(s_i, s_{i+1})$:
\begin{equation}
    p_i = 1 - \frac{|J(s_i, s_{i+1}) - 0.25|}{0.75}, \qquad
    \rcf_\text{prog} = \max\!\left(0,\,
    \frac{1}{n-1}\sum_{i=1}^{n-1} p_i\right)
\end{equation}
Target overlap 0.25 reflects the empirical distribution of consecutive-step Jaccard similarities in the reference corpus ($\mu{=}0.24$, $\sigma{=}0.07$, $n{=}500$).
 
\paragraph{Convergence.}
$\rcf_\text{conv} = J(s_n, H)$, where $H$ is the generated hypothesis.
 
\paragraph{Non-redundancy.}
\begin{equation}
    \rcf_\text{nr} = \max\!\left(0,\,
    1 - \frac{2}{n(n-1)}\sum_{i<j} J(s_i, s_j)\right)
\end{equation}
 
\paragraph{Density.}
Let $\bar{w} = \frac{1}{n}\sum_k |s_k|_\text{words}$.
\begin{align}
    \sigma_\text{low}  &= \tfrac{1}{1+e^{-0.05(\bar{w}-40)}}, \quad
    \sigma_\text{high} = \tfrac{1}{1+e^{-0.02(\bar{w}-150)}} \\
    \rcf_\text{den}    &= \max\!\left(0,\,
    \min(\sigma_\text{low},\, 1 - \sigma_\text{high})\right)
\end{align}
The double-sigmoid peaks at $\bar{w}{\approx}80$ words/step
and decays toward 0 for $\bar{w}{<}20$ or $\bar{w}{>}200$.

\subsection{Hypothesis-Problem Alignment (\hpa{})}
\label{app:hpa}
 
Let $P$, $H$, $F$, $I$, $M$, $T$ denote the problem statement, hypothesis, failure mode, intervention, mechanism, and target property fields respectively.
 
\paragraph{Causal language sub-score.}
$\mathcal{C}$ is a set of 13 causal connector patterns: \{thereby, thus, leading to, resulting in, which enables, by $\ldots$ increas-, enhancing, reducing, improving, will $\ldots$ enable, provides, allows, facilitates\}.
\begin{equation}
    \hpa_\text{causal} =
    \min\!\left(1,\,\frac{|\{c \in \mathcal{C} : c \text{ matches } H\}|}{4}\right)
\end{equation}
The denominator 4 is calibrated so that $\geq 4$ distinct
causal connectors in a 2--3 sentence hypothesis indicates
saturated causal expressivity (${\geq}95$th percentile of
reference corpus).
 
\paragraph{Fallbacks.}
If $F = \varnothing$: $\hpa_\text{fail} = 0.8 \cdot J(P,H)$.
If $T = \varnothing$: $\hpa_\text{tgt} = 0.5$.
If $I = \varnothing$ or $M = \varnothing$:
$\hpa_\text{coh} = 0.3$.

\subsection{Mechanistic Specificity Index (\msi{})}
\label{app:msi}
 
Let $C = H \oplus M \oplus R$ (concatenation of hypothesis, mechanism, reasoning process).
 
\paragraph{Vocabulary tiers.}
Three curated vocabulary sets, all entries lowercased and matched as substrings in $C$:
\begin{itemize}
    \item $\mathcal{K}_H$ (44 terms, weight 3): electrochemical mechanism terms, e.g.Butler-Volmer,
    solid electrolyte interphase, tortuosity, operando, Coulombic efficiency, Jahn-Teller, lattice parameter, formation energy.
    \item $\mathcal{K}_M$ (23 terms, weight 1.5):  materials/component terms, e.g.\ doping, porosity, capacity fade, volume expansion, SEI, grain boundary.
    \item $\mathcal{K}_L$ (13 terms, weight 0.5): vague filler terms, e.g.\ improve, enhance, novel, promising, efficient.
\end{itemize}
\begin{equation}
    \msi_\text{vocab} =
    \min\!\left(1,\,\frac{3h + 1.5m + 0.5\ell}
    {3|\mathcal{K}_H| + 1.5|\mathcal{K}_M| + 0.5|\mathcal{K}_L|}
    \times 3\right)
\end{equation}
The $\times 3$ normalization ensures the score spans $[0,1]$ for typical scientific texts; without it, the denominator is too large for any single hypothesis to saturate.
 
\paragraph{Quantitative grounding.}
$\mathcal{Q}$: 10 regex patterns matching physical quantities with units, including nm/\textmu m/mm, $^\circ$C/K, eV/kJ, mAh/Wh, mol, V/mV, mA/cm$^2$, \%, mg/g/cm$^3$, and scientific notation ($a{\times}10^b$).
Saturates at 5 distinct matches.
 
\paragraph{Characterization anchoring.}
$\mathcal{X}$: 23 techniques---XRD, TEM, SEM, XPS, EIS, NMR, Raman, FTIR, DFT, AIMD, synchrotron, neutron diffraction, operando, cryo-TEM, SAXS, WAXS, DSC, TGA, GITT, PITT, cyclic voltammetry, galvanostatic, impedance spectroscopy. Saturates at 4 distinct matches.
 
\paragraph{Mechanism depth.}
$\msi_\text{depth} = \min(1,\, |M|_w / (3|H|_w))$. Saturates when the mechanism rationale is $\geq 3{\times}$ the word count of the hypothesis, consistent with the median ratio of 3.2 in the reference corpus.

\subsection{Scientific Novelty Score (\sns{})}
\label{app:sns}
 
\paragraph{TF-IDF construction.}
For corpus $\{H_1,\ldots,H_N\}$, vocabulary $V{=}2000$
(top terms by document frequency, 3-character minimum):
\begin{equation}
    X_{ij} = \frac{c_{ij}}{\sum_k c_{ik}}
    \times \left(\log\frac{N+1}{df_j+1} + 1\right)
\end{equation}
Rows are $\ell_2$-normalised. The cosine similarity matrix
is $\mathbf{S} = \mathbf{X}\mathbf{X}^\top$ with
$S_{ii} = -1$ to exclude self-similarity.
 
\paragraph{Corpus novelty.}
\begin{equation}
    \sns_\text{corp}(i) =
    1 - \frac{1}{5}\sum_{k=1}^{5} S_{i,\pi_i(k)}
\end{equation}
where $\pi_i(k)$ is the $k$-th most similar hypothesis index. Top-5 mean (not max) is used for robustness against a single near-duplicate.
 
\paragraph{Within-system novelty.}
For each battery system group $\mathcal{B}$, the same pipeline is run on the submatrix $\mathbf{X}_\mathcal{B}$ using top-3 neighbors (smaller groups have fewer comparators).
 
\paragraph{Cross-domain bonus.}
$\mathcal{D}$: 28 terms from adjacent fields biomimetic, aerogel, MOF, COF, zeolite, metamaterial, topology, fractal, quantum, plasma, textile, machine learning, neural network, gasification, wood-derived, bio-inspired, biomass, silk, cellulose, chitin, and 8 others. Saturates at 2 distinct hits.
 
 
\subsection{Intervention Plausibility (\ip{})}
\label{app:ip}
 
\paragraph{Material compatibility.}
System-to-material keyword mapping (6 families):
\begin{center}
\small
\begin{tabular}{ll}
\toprule
System & Expected constituent keywords \\
\midrule
NMC & li, ni, mn, co, oxide, layered \\
LFP & fe, phosphate, olivine, iron \\
NCA & ni, co, al, layered \\
Silicon & si, silicon, expansion, volume \\
Solid-state & solid, ceramic, sulfide, oxide, garnet \\
Li metal & li metal, dendrite, plating \\
\bottomrule
\end{tabular}
\end{center}
$\ip_\text{mat} = \min(1,\, 0.5 + 0.1 \times \text{hits})$.
Defaults to 0.5 for unrecognised systems.
 
\paragraph{Scalability.}
$\mathcal{S}^+$: 14 terms (scalable, cost-effective, low-cost, roll-to-roll, industrial, commercializ-, mass produc-, pilot, kg-scale, ton-scale, solution process, spray coat, simple, facile). $\mathcal{S}^-$: 4 terms (atomic layer deposition [unqualified], CVD without scalability qualifier, extremely expensive, requires ultra-high vacuum).
\begin{equation}
    \ip_\text{scale} =
    \max\!\left(0,\,\min\!\left(1,\,
    0.3 + 0.15|\mathcal{S}^+ \cap C|
        - 0.2|\mathcal{S}^- \cap C|\right)\right)
\end{equation}
 
\paragraph{Evidence grounding.}
Mapped from \texttt{evidence\_strength} field: strong/high ${\to}1.0$; moderate ${\to}0.65$; theoretical ${\to}0.50$; preliminary ${\to}0.40$; unknown ${\to}0.35$; weak ${\to}0.25$.
 
\paragraph{Outcome specificity.}
$\ip_\text{out} = 0.8$ if $O$ matches a numeric quantity with unit (regex: \texttt{\textbackslash d+\%|\textbackslash d+\,mAh|\textbackslash d+\,V}, etc.);
$0.6$ if $|O|_w > 10$; else $0.3$.
 
 
\subsection{Problem Decomposition Quality (\pdq{})}
\label{app:pdq}
 
\paragraph{Root-cause compression.}
Let $r = |P_\text{core}|_w / |P|_w$:
\begin{equation}
    \pdq_\text{core} =
    \begin{cases}
        1.0 & 0.15 \leq r \leq 0.50 \\
        0.7 & 0.05 \leq r < 0.15
              \;\text{or}\; 0.50 < r \leq 0.80 \\
        0.4 & r > 0.80 \\
        0.3 & r < 0.05
    \end{cases}
\end{equation}
 
\paragraph{Failure mode specificity.}
$f_\text{spec}$: count of $\mathcal{K}_H$ terms in $F$.
$f_\text{vague}$: count of \{poor, bad, issue, problem,
challenge, difficulty, limitation\} in $F$.
$\pdq_\text{fail} = \max(0,\,\min(1,\, 0.3
+ 0.2 f_\text{spec} - 0.1 f_\text{vague}))$.
 
\paragraph{Abstraction consistency.}
$\pdq_\text{abs} = \max(0.2,\, J(P_\text{broad}, P_\text{fine}))$. Floor of 0.2 accounts for valid IS-A relationships without shared surface tokens (e.g., ``thermal etching'' specialises
``mass transport'').

\section{Metric Validity and Remaining Limitations}
\label{app:metric_validity}

The six metrics in \bench{} are intended as transparent operational proxies for domain-relevant aspects of hypothesis quality, not as ground-truth measures of scientific discovery. We therefore assess them through converging validity checks rather than a single definitive test. \ref{sec:metric_tradeoffs} shows that the dimensions separate different system behaviors: for example, high \rcf{} does not imply high \msi{}. Tests in \ref{sec:error_taxonomy} checks whether low-scoring regions correspond to interpretable scientific failure modes. Comparisons with BLEU, ROUGE-L, BERTScore, and embedding cosine similarity test whether \cbs{} captures signal beyond conventional reference matching. \ref{sec:judge_validation} provides noisy external evidence, while gaming stress test in \ref{sec:gaming_validation} checks whether the aggregate remains stable under superficial adversarial manipulation. Finally, \ref{app:weight_ablation} shows that the main ranking is stable under reasonable perturbations of the \cbs{} weights. These checks do not prove that the equations perfectly reflect expert scientific judgment, but they support the claim that the metric suite behaves consistently with the intended constructs.

Several limitations remain. First, \ip{} uses evidence strength as one subcomponent, but evidence strength is extracted from the source paper and may be noisy when a paper provides incomplete or ambiguous experimental support. Second, instance construction relies on automated extraction from papers, with expert checking of the schema and a subset of instances rather than full manual annotation of all examples. Third, retrieval and direct prompting are nearly tied in overall \cbs{}, suggesting that generic retrieval does not automatically improve mechanistic grounding unless the retrieved context contains the right causal details. Fourth, the reference set has lower \hpa{} than some generated systems because the literature-derived hypothesis can be narrower, more mechanism-focused, or less lexically aligned with the extracted problem statement than a model-generated answer that mirrors the problem text. We view these as important directions for future validation, especially through larger expert annotation studies and downstream checks against simulation or experimental outcomes.
 
\paragraph{Component granularity.}
$\mathcal{G}$: 14 specific component terms (cathode, anode, sei, cei, interface, grain boundary, particle, electrode, binder, active material, current collector, separator, nlp, carbon black).
$\pdq_\text{comp} = \min(1,\, 0.4 \cdot \mathbb{1}[|G|_w>1]
+ 0.4 \cdot |\mathcal{G} \cap G|)$.

\section{Pairwise LLM-Judge Prompt}
\label{app:judge_prompt}

\begin{tcolorbox}[
  colback=green!5!white,
  colframe=green!45!black,
  fonttitle=\bfseries\small,
  title=LLM Judge Prompt,
  boxrule=0.6pt,
  left=4pt, right=4pt, top=3pt, bottom=3pt,
]
\textbf{System:} You are an expert battery materials scientist evaluating two AI-generated research hypotheses.

\smallskip
\textbf{Input:} \texttt{PROBLEM}, \texttt{BATTERY\_SYSTEM}, \texttt{FAILURE\_MODE}, \texttt{HYPOTHESIS A}, \texttt{HYPOTHESIS B}

\smallskip
Evaluate which hypothesis is better on three criteria:\\
\textbf{1.} \textsc{Problem\_Addressed} --- which better addresses the stated failure mode?\\
\textbf{2.} \textsc{Mechanistic\_Depth} --- which is more mechanistically grounded?\\
\textbf{3.} \textsc{Scientific\_Utility} --- which would be more useful to a researcher?

\smallskip
\textit{Respond only with JSON:}
{\ttfamily\small \{"problem\_addressed": "A|B", "mechanistic\_depth": "A|B", "scientific\_utility": "A|B", "overall": "A|B", "confidence": "high|medium|low"\}}

\smallskip
Applied twice per pair with swapped order (A$\to$B, then B$\to$A) to control for position bias.
\end{tcolorbox}


\section{Scoring Example}
\label{app:worked_example}

\begin{tcolorbox}[
  enhanced,
  breakable,
  colback=orange!4!white,
  colframe=orange!55!black,
  coltitle=white,
  boxrule=0.7pt,
  arc=2pt,
  left=5pt,
  right=5pt,
  top=5pt,
  bottom=5pt,
  fonttitle=\bfseries\small,
  title={Scoring example: reference-free evaluation}
]
\small
\textbf{Problem.}
A disordered-rocksalt cathode has poor Li$^+$ transport, low electronic conductivity, and capacity fade at elevated C-rates.

\medskip
\textbf{Three hypotheses.}

\textbf{Reference:} Fluorination, particle-size reduction, and carbon coating improve Li$^+$ percolation, shorten diffusion paths, improve electronic transport, and suppress irreversible oxygen loss.\\

\textbf{Fluent/vague:} Optimizing composition and microstructure can improve Li-ion transport and conductivity, leading to better rate capability and cycling stability.\\

\textbf{Alternative:} Mo$^{6+}$/Nb$^{5+}$ co-substitution stabilizes 0-TM Li diffusion channels, lowers Li migration barriers, improves high-rate capacity, and suppresses Mn-dissolution-driven fade.

\medskip
\begin{center}
\setlength{\tabcolsep}{4pt}
\renewcommand{\arraystretch}{0.95}
\begin{tabular}{lccc}
\toprule
\textbf{Metric} & \textbf{Reference} & \textbf{Fluent/vague} & \textbf{Alternative} \\
\midrule
\rcf{} & 0.71 & 0.74 & 0.69 \\
\hpa{} & 0.34 & 0.22 & 0.31 \\
\msi{} & 0.42 & 0.03 & 0.51 \\
\sns{} & 0.78 & 0.81 & 0.84 \\
\ip{}  & 0.82 & 0.51 & 0.79 \\
\pdq{} & 0.61 & 0.52 & 0.61 \\
\midrule
\cbs{} & 0.608 & 0.403 & 0.612 \\
\bottomrule
\end{tabular}
\end{center}

\textbf{Takeaway.}
The fluent/vague hypothesis has the highest \rcf{} but very low \msi{}, showing that coherent reasoning is not the same as mechanistic specificity. The alternative hypothesis scores slightly above the reference even though it does not match the paper's intervention, illustrating that \bench{} rewards scientific quality rather than overlap with a single reference answer.
\end{tcolorbox}

\section{Design Rationale: Programmatic vs.\ LLM-as-Judge}
\label{app:design_rationale}

We deliberately avoid LLM-as-judge for the six core metrics. This choice is motivated by three considerations:

\textbf{(1) Reproducibility.}

Programmatic metrics produce identical scores given identical inputs, independent of model version, API availability, or sampling stochasticity. LLM judges are known to exhibit significant score variance across calls and model versions~\citep{2025arXiv250818076C}.

\textbf{(2) Interpretability.}

Because the benchmark is intended to diagnose failure modes, we prefer metrics whose inputs and behavior can be inspected directly. Programmatic dimensions make it easier to identify why a system scored poorly, for example due to weak mechanism, weak problem alignment, or poor decomposition, rather than only observing a single opaque judgment.

\textbf{(3) Cost and accessibility.}

Running \bench{} on 2{,}645 hypotheses with an LLM judge would cost approximately \$150-\$400 per evaluation run at current API prices, making re-evaluation on new systems prohibitively expensive. The programmatic pipeline runs in less than 90 seconds on a standard CPU node. An optional LLM-judge module (\textsc{cge}: Co-Scientist Generation Evaluator) is provided for supplementary deep evaluation of scientific soundness and falsifiability, and is excluded from the main \cbs{} computation.

\clearpage


\end{document}